\documentstyle[preprint,aps]{revtex}
\begin{document}
\title{Blue Perturbation Spectra from Hybrid Inflation
 with Canonical Supergravity}
\author{C. Panagiotakopoulos}
\address{Physics Division\\
School of Technology\\
University of Thessaloniki\\
Thessaloniki, Greece}
\maketitle

\begin{abstract}
We construct a hybrid inflationary model associated with the superheavy
scale $M_X \sim 10^{16}$ GeV of supersymmetric grand unified theories in
which the inflaton potential is provided entirely by canonical supergravity.
We find that the spectrum of adiabatic density perturbations is
characterized by a strongly varying spectral index which is considerably
larger than unity. Moreover, the total number of e-foldings is very limited.
Implications of our analysis for other hybrid inflationary scenarios are
briefly discussed.
\end{abstract}

\newpage

The majority of successful inflationary scenarios \cite{linde90} invoke a
very weakly coupled gauge singlet scalar field known as inflaton in order to
account for the smallness of the observed temperature fluctuations $\frac{%
\Delta T}{T}$ in the cosmic background radiation. Recently, Linde \cite
{linde94} proposed a new inflationary model which looks as a hybrid of
chaotic inflation with a quadratic potential for the gauge singlet inflaton
and the usual theory of spontaneous symmetry breaking involving a possibly
gauge non-singlet field. During inflation the non-inflaton field is trapped
in a false vacuum state and the universe is dominated by the false vacuum
energy density. Inflation ends with a phase transition when the non-inflaton
field rolls very rapidly to its true vacuum state (``waterfall''). The
beauty of this hybrid model lies in the fact that it does not use small
coupling constants in order to produce the observed temperature fluctuations
and that it reconnects inflation with phase transitions in grand unified
theories (GUTs). Although the original hybrid model is non-supersymmetric,
there is a direct adjustment to supersymmetric (SUSY) models \cite{cope}
with the most commonly used superpotential for symmetry breaking, an
inflaton mass of the order of $1\ TeV$, the SUSY breaking scale, and an
intermediate scale $(\sim 10^{11}-10^{12}\>GeV)$ of symmetry breaking. An
additional motivation for a SUSY hybrid inflationary scenario is the
possibility offered by supersymmetry to naturally forbid large
self-couplings of the inflaton through R-symmetries \cite{dvali}. In
connection with supersymmetry it is very desirable to associate hybrid
inflation with symmetry breaking scales $\sim 10^{16}\>GeV$ consistent with
the unification of the gauge couplings of the minimal supersymmetric
standard model (MSSM) which is favored by LEP data. However, the electroweak
mass of the inflaton provided by SUSY breaking is too weak to account for
the correct value of $\frac{\Delta T}{T}$ and an appropriate potential for
the inflaton has to be found \cite{dvali}. In particular, a variation of
hybrid inflation, smooth hybrid inflation \cite{laz}, in which the phase
transition takes place gradually during inflation, has been invented to
successfully address this issue. As soon as one replaces global by local
supersymmetry it is well-known that the potential becomes very steep and
inflation becomes, in general, impossible. This is, to a large extent, due
to the generation of a mass for the inflaton which is larger than the Hubble
constant $H$. For the simple superpotentials used so far in SUSY bybrid
inflation such a mass for the inflaton is not generated provided the
canonical form of the K$\ddot{a}$hler potential of $N=1$ supergravity is
employed \cite{cope}. However, even in these cases one has to question the
extent to which canonical supergravity affects a successful global SUSY
inflationary scenario, especially if during inflation the inflaton takes
values close to the supergravity scale $M_{P}/\sqrt{8\pi }\simeq
2.4355\times 10^{18}\>GeV\>\>(M_{P}\simeq 1.221\times 10^{19}\>GeV$ is the
Planck mass). Intermediate scale models are not expected to be seriously
affected by the additional terms generated by canonical supergravity whereas
the same does not apply to models in which inflation is associated with the
superheavy scale $M_{X}\sim 10^{16}\>GeV$ of SUSY GUTs.

A logically distinct possibility is that during inflation supergravity
dominates the inflaton potential instead of simply being a perturbation. Our
purpose in the present paper is to investigate the above possibility and
attempt to construct a hybrid inflationary model associated with the
superheavy scale of SUSY GUTs in which the inflaton potential is provided
entirely by the terms generated when global supersymmetry is replaced by
canonical supergravity. We will see that the most natural implementation of
this idea is in the context of SUSY GUTs based on semi-simple gauge groups
of rank six. An interesting property of supergravity dominated hybrid
inflation is that the spectral index of the adiabatic density perturbations
is considerably larger than 1 (blue primordial spectra\cite{mol}) and
strongly varying.

We consider a SUSY GUT based on a (semi-simple) gauge group $G$ of rank $%
\geq 5$. $G$ breaks spontaneously directly to the standard model (SM) gauge
group $G_{S}$ $\equiv SU(3)_{c}\times SU(2)_{L}\times U(1)_{Y}$ at a scale $%
M_{X}\sim 10^{16}\>GeV$. The symmetry breaking of $G$ to $G_{S}$ is obtained
through a superpotential which includes the terms 
\begin{equation}
W=S(-\mu ^{2}+\lambda \Phi \bar{\Phi}).
\end{equation}
Here $\Phi ,\bar{\Phi}$ is a conjugate pair of left-handed SM singlet
superfields which belong to non-trivial representations of $G$ and reduce
its rank by their vacuum expectation values (vevs), $S$ is a gauge singlet
left-handed superfield, $\mu $ is a superheavy mass scale related to $M_{X}$
and $\lambda $ a real and positive coupling constant. The superpotential
terms in eq. (1) are the dominant couplings involving the superfields $S$, $%
\Phi $, $\bar{\Phi}$ which are consistent with a continuous R-symmetry under
which $W\to e^{i\gamma }W$, $S\to e^{i\gamma }S$, $\Phi \to \Phi $ and $\bar{%
\Phi}\to \bar{\Phi}$. Moreover, we assume that the presence of other $SM$
singlets in the theory does not affect the superpotential in eq. (1). The
potential obtained from $W$, in the supersymmetric limit, is 
\begin{equation}
V=\mid -\mu ^{2}+\lambda \Phi \bar{\Phi}\mid ^{2}+\mid \lambda S\mid
^{2}(\mid \Phi \mid ^{2}+\mid \bar{\Phi}\mid ^{2})+D-terms,
\end{equation}
where the scalar components of the superfields are denoted by the same
symbols as the corresponding superfields. The SUSY vacuum 
\begin{equation}
<S>=0,<\Phi ><\bar{\Phi}>=\mu ^{2}/\lambda ,\>\>\mid <\Phi >\mid =\mid <\bar{%
\Phi}>\mid
\end{equation}
lies on the D-flat direction $\Phi =\bar{\Phi}^{*}$. By appropriate gauge
and R-trasformations on this D-flat direction we can bring the complex $S$, $%
\Phi $, $\bar{\Phi}$ fields on the real axis, i.e. $S\equiv \frac{1}{\sqrt{2}%
}\sigma $, $\Phi =\bar{\Phi}\equiv \frac{1}{2}\phi $, where $\sigma $ and $%
\phi $ are real scalar fields. The potential in eq. (2) then becomes 
\begin{equation}
V(\phi ,\sigma )=(-\mu ^{2}+\frac{1}{4}\lambda \phi ^{2})^{2}+\frac{1}{4}%
\lambda ^{2}\sigma ^{2}\phi ^{2}
\end{equation}
and the supersymmetric vacuum corresponds to $\mid <\frac{\phi }{2}>\mid =%
\frac{\mu }{\sqrt{\lambda }}=\frac{M_{X}}{g}$ and $<\sigma >=0$, where $%
M_{X} $ is the mass acquired by the gauge bosons and $g$ is the gauge
coupling constant. For any fixed value of $\sigma >\sigma _{c}$, where $%
\sigma _{c}=\sqrt{2}\mu /\sqrt{\lambda }=\sqrt{2}\mid <\frac{\phi }{2}>\mid $%
, $V$ as a function of $\phi $ has a minimum lying at $\phi =0$. The value
of $V$ at this minimum for every value of $\sigma >\sigma _{c}$ is $\mu ^{4}$%
.

Adding to $V$ a mass-squared term for $\sigma $ we obtain Linde's potential
with the only difference that in the SUSY model the critical value $\sigma
_{c}$ of $\sigma $, below which the minimum at $\phi =0$ disappears, becomes
very closely connected with the vev of $\phi $. When $\sigma >\sigma _{c}$
the universe is dominated by the false vacuum energy density $\mu ^{4}$ and
expands quasi-exponentially. When $\sigma $ falls below $\sigma _{c}$ the
mass-squared term of $\phi $ becomes negative, the false vacuum state at $%
\phi =0$ becomes unstable and $\phi $ rolls rapidly to its true vacuum
thereby terminating inflation.

Let us now replace global supersymmetry by $N=1$ canonical supergravity.
From now on we will use the units in which $\frac{M_{P}}{\sqrt{8\pi }}=1$.
Then, the potential $V(\phi ,\sigma )$ becomes 
\begin{equation}
V(\phi ,\sigma )=[(-\mu ^{2}+\frac{1}{4}\lambda \phi ^{2})^{2}(1-\frac{%
\sigma ^{2}}{2}+\frac{\sigma ^{4}}{4})+\frac{1}{4}\lambda ^{2}\sigma
^{2}\phi ^{2}(1-\frac{\mu ^{2}}{\lambda }+\frac{1}{4}\phi ^{2})^{2}]%
\displaystyle{e^{\frac{1}{2}(\sigma ^{2}+\phi ^{2})}}.
\end{equation}
$V$ still has a minimum with $V=0$ at $\mid \frac{\phi }{2}\mid =\frac{\mu }{%
\sqrt{\lambda }}$ and $\sigma =0$ and a critical value $\sigma _{c}$ of $%
\sigma $ which remains essentially unaltered. The important difference lies
in the expression of $V(\sigma )$ for $\sigma >\sigma _{c}$ and $\phi =0$ 
\begin{equation}
V(\sigma )=\mu ^{4}(1-\frac{\sigma ^{2}}{2}+\frac{\sigma ^{4}}{4})%
\displaystyle{e^{\frac{\sigma ^{2}}{2}}},
\end{equation}
which now has a non-zero derivative $V^{\prime }(\sigma )$ with respect to $%
\sigma $ 
\begin{equation}
V^{\prime }(\sigma )=\frac{1}{2}\mu ^{4}\sigma ^{3}(1+\frac{\sigma ^{2}}{2})%
\displaystyle{e^{\frac{\sigma ^{2}}{2}}}.
\end{equation}

Expanding $V(\sigma )$ in powers of $\sigma ^{2}$ and keeping the first
non-constant term only we obtain 
\begin{equation}
V(\sigma )\simeq \mu ^{4}+\frac{1}{8}\mu ^{4}\sigma ^{4}\qquad (\sigma
^{2}<<1).
\end{equation}
We see that no mass-squared term for $\sigma $ is generated \cite{cope} and
that for $\sigma ^{2}<<1$ the model resembles the original hybrid
inflationary model with a quartic, instead of quadratic, inflaton potential
in which the quartic coupling takes naturally the very small value $\frac{1}{%
2}\mu ^{4}$. The number of e-foldings $\Delta N(\sigma _{in},\sigma _{f})$
for the time period that $\sigma $ varies between the values $\sigma _{in}$
and $\sigma _{f}\>\ (\sigma _{in}>\sigma _{f})$ is given, in our
approximation, by 
\begin{equation}
\Delta N(\sigma _{in},\sigma _{f})=-\int_{\sigma _{in}}^{\sigma {_{f}}}\frac{%
V}{V^{\prime }}d\sigma =\sigma _{f}^{-2}-\sigma _{in}^{-2}.
\end{equation}
Also the ratio $(\frac{\Delta T}{T})_{T}^{2}/(\frac{\Delta T}{T}%
)_{S}^{2}\sim \sigma _{H}^{6}<<1$, where $(\frac{\Delta T}{T})_{T}$ and $(%
\frac{\Delta T}{T})_{S}$ are the tensor and scalar components of the
quadrupole anisotropy $\frac{\Delta T}{T}$ respectively and $\sigma _{H}$ is
the value that the inflaton field had when the scale $\ell _{H}$,
corresponding to the present horizon, crossed outside the inflationary
horizon. Therefore, we can safely ignore $(\frac{\Delta T}{T})_{T}$ and
obtain \cite{lid} 
\begin{equation}
\frac{\Delta T}{T}\simeq \frac{1}{4\pi \sqrt{45}}(\frac{V^{3/2}}{V^{\prime }}%
)_{\sigma {_{H}}}=\frac{1}{2\pi \sqrt{45}}\frac{\mu ^{2}}{\sigma _{H}^{3}}%
\qquad (\sigma _{H}^{2}<<1).
\end{equation}

Using the above approximate expressions we are going to investigate the
possibility that $V(\sigma )$ is the inflaton potential. Let us assume first
that $\sigma _{H}\simeq \sigma _{c}$. From eq. (9) this happens only if $%
\sigma _{H}^{2}\simeq \sigma _{c}^{2}<<N_{H}^{-1}$, where $N_{H}\>\>$ $%
(\simeq 50-60)$ is the number of e-foldings for the time period that $\sigma 
$ varies between $\sigma _{H}$ and $\sigma _{c}$. Then, for $\sigma
_{H}\simeq \sigma _{c}$ and $\frac{\Delta T}{T}\simeq 6.6\times 10^{-6}$,
eq. (10) gives 
\begin{equation}
\mu \simeq 0.0281(\frac{\sigma _{c}}{\sqrt{2}\mid <\frac{\phi }{2}>\mid })^{%
\frac{3}{2}}\mid <\frac{\phi }{2}>\mid ^{\frac{3}{2}}\qquad (\sigma
_{c}^{2}<<N_{H}^{-1}).
\end{equation}
If we assume the relation $\sigma _{c}\simeq \sqrt{2}\mid <\frac{\phi }{2}%
>\mid $, which holds in our simple model, and use the MSSM values $%
M_{X}=2\times 10^{16}$ GeV, $g=0.7$ to calculate $\mid <\frac{\phi }{2}>\mid
=\frac{M_{X}}{g}\simeq 0.01173$, we obtain from eq. (11) $\mu \simeq
3.57\times 10^{-5}$ and $\lambda \simeq 9.26\times 10^{-6}$. We see that,
for scales as large as the scale implied by the MSSM, $\mu <<\mid <\frac{%
\phi }{2}>\mid $ and $\lambda <<1$. For smaller scales the situation becomes
even worse because $\frac{\mu }{\mid <\frac{\phi }{2}>\mid }$ $\sim \mid <%
\frac{\phi }{2}>\mid ^{\frac{1}{2}}$. The only essentially different
possibility is that $\sigma _{H}^{2}>>N_{H}^{-1}\simeq \sigma _{c}^{2},$
i.e. 
\begin{equation}
\mid <\frac{\phi }{2}>\mid \simeq (2N_{H})^{-\frac{1}{2}}(\frac{\sqrt{2}\mid
<\frac{\phi }{2}>\mid }{\sigma _{c}})\qquad (\sigma _{c}\simeq N_{H}^{-\frac{%
1}{2}}).
\end{equation}
Again, if the relation $\sigma _{c}\simeq \sqrt{2}\mid <\frac{\phi }{2}>\mid 
$ holds, this possibility is ruled out for the scale of MSSM. However, a
scale $\sim $ $10^{17}$ $GeV$, implied by the relation $N_{H}^{-\frac{1}{2}%
}\simeq $ $\sigma _{c}\simeq \sqrt{2}\mid <\frac{\phi }{2}>\mid $, should
not be excluded in SUSY GUTs with a spectrum of states different from the
spectrum of MSSM. Our subsequent discussion of inflation applies to this
case as well.

The above arguments lead to the conclusion that if in the context of MSSM we
insist in avoiding $\mu \,\ll \,\mid <\frac{\phi }{2}>\mid $, the choice of $%
V(\sigma )$ as the inflaton potential leads to acceptable values of $\frac{%
\Delta T}{T}$ only if the relation $\sigma _{c}\simeq \sqrt{2}\mid <\frac{%
\phi }{2}>\mid $ of the simplest supersymmetric model is violated in order
to allow $\sigma _{c}>>\mid <\frac{\phi }{2}>\mid $. This will be achieved
by introducing in the discussion a second gauge non-singlet field acquiring
a large vev. As a consequence the minimum rank of the gauge group $G$ must
be extended from five to six. We do not regard this extention as a serious
restriction, but rather as a different treatment of existing fields in
realistic models, since most semi-simple gauge groups used in model building
are rank-six subgroups of $E_{6}$.

The superpotential responsible for the breaking of $G$ to $G_{S}$ includes
now the terms 
\begin{equation}
W=\tilde{S}(-\mu _{1}^{2}+\tilde{\lambda}_{1}\Phi _{1}\bar{\Phi}_{1}+\tilde{%
\lambda}_{2}\Phi _{2}\bar{\Phi}_{2})+\tilde{S}^{\prime }(-\mu _{2}^{2}+%
\tilde{\lambda}_{3}\Phi _{1}\bar{\Phi}_{1}+\tilde{\lambda}_{4}\Phi _{2}\bar{%
\Phi}_{2}).
\end{equation}
We have two conjugate pairs $\Phi _{1},\bar{\Phi}_{1}$ and $\Phi _{2},\bar{%
\Phi}_{2}$ of left-handed $SM$ singlet superfields which belong to
non-trivial representations of $G$ and whose vevs reduce its rank by two
units, $\tilde{S},\tilde{S}^{\prime }$ are gauge singlet left-handed
superfields, $\mu _{1},\mu _{2}$ are superheavy masses $\sim M_{X}$ and $%
\tilde{\lambda}_{1},\tilde{\lambda}_{2},\tilde{\lambda}_{3},\tilde{\lambda}%
_{4}$ real coupling constants. Under the continuous R-symmetry $W\to
e^{i\gamma }W$, $\tilde{S}\to e^{i\gamma }\tilde{S}$, $\tilde{S}^{\prime
}\to e^{i\gamma }\tilde{S}^{\prime }$ with the remaining superfields
transforming trivially. Let us define $\mu ^{2}\equiv (\mu _{1}^{4}+\mu
_{2}^{4})^{1/2},cos\theta \equiv \mu _{1}^{2}/\mu ^{2}$, $sin\theta \equiv
\mu _{2}^{2}/\mu ^{2},S\equiv cos\theta \tilde{S}+sin\theta \tilde{S}%
^{\prime },S^{\prime }\equiv -sin\theta \tilde{S}+cos\theta \tilde{S}%
^{\prime },\lambda _{1}\equiv cos\theta \tilde{\lambda}_{1}+sin\theta \tilde{%
\lambda}_{3},\lambda _{2}\equiv cos\theta \tilde{\lambda}_{2}+sin\theta 
\tilde{\lambda}_{4},\lambda _{3}\equiv -sin\theta \tilde{\lambda}%
_{1}+cos\theta \tilde{\lambda}_{3}$ and $\lambda _{4}\equiv sin\theta \tilde{%
\lambda}_{2}-cos\theta \tilde{\lambda}_{4}$. Then, $W$ becomes 
\begin{equation}
W=S(-\mu ^{2}+\lambda _{1}\Phi _{1}\bar{\Phi}_{1}+\lambda _{2}\Phi _{2}\bar{%
\Phi}_{2})+S^{\prime }(\lambda _{3}\Phi _{1}\bar{\Phi}_{1}-\lambda _{4}\Phi
_{2}\bar{\Phi}_{2}).
\end{equation}
Along the D-flat directions of the potential the symmetries of W allow us to
define real scalar fields $\phi _{1},\phi _{2},\sigma ,\sigma _{1},\sigma
_{2}$ through the relations $\Phi _{1}=\bar{\Phi}_{1}\equiv \frac{1}{2}\phi
_{1},\Phi _{2}=\bar{\Phi}_{2}\equiv \frac{1}{2}\phi _{2},S\equiv \frac{1}{%
\sqrt{2}}\sigma $ and $S^{\prime }\equiv \frac{1}{\sqrt{2}}(\sigma
_{1}+i\sigma _{2})$. The potential obtained from $W$ is then given by 
\[
V(\phi _{1},\phi _{2},\sigma ,\sigma _{1},\sigma _{2})=(-\mu ^{2}+\frac{1}{4}%
\lambda _{1}\phi _{1}^{2}+\frac{1}{4}\lambda _{2}\phi _{2}^{2})^{2}+\frac{1}{%
16}(\lambda _{3}\phi _{1}^{2}-\lambda _{4}\phi _{2}^{2})^{2}\qquad \qquad
\qquad \qquad 
\]
\begin{equation}
\text{ }\qquad \qquad \qquad \quad +\frac{1}{4}[(\lambda _{1}\sigma +\lambda
_{3}\sigma _{1})^{2}+\lambda _{3}^{2}\sigma _{2}^{2}]\phi _{1}^{2}+\frac{1}{4%
}[(\lambda _{2}\sigma -\lambda _{4}\sigma _{1})^{2}+\lambda _{4}^{2}\sigma
_{2}^{2}]\phi _{2}^{2}.
\end{equation}
We assume that $\lambda _{1},\lambda _{2},\lambda _{3},\lambda _{4}$ are
real and positive couplings which satisfy the constraints $\lambda
_{2}>>\lambda _{1}>\lambda _{3}=\lambda _{4}$. Then $V$ has a SUSY minimum
at $\mid \frac{\phi _{1}}{2}\mid =\mid \frac{\phi _{2}}{2}\mid =\mu /\sqrt{%
\lambda _{1}+\lambda _{2}}\simeq \mu /\sqrt{\lambda _{2}}$ and $\sigma
=\sigma _{1}=\sigma _{2}=0$. Moreover, for fixed $\sigma >>\sigma
_{1},\,\sigma _{2},\,V$ as a function of $\phi _{1}$ and $\phi _{2}$ has a
minimum at $\phi _{1}=\phi _{2}=0$ provided $\sigma >\sigma _{c}\simeq \sqrt{%
2}\mu /\sqrt{\lambda _{1}}$. If we set $\phi _{1}=\phi _{2}=0$ in $V$ we
find again that $V=\mu ^{4}$. Once more there is not a potential neither for 
$\sigma $ nor for $\sigma _{1}$ and $\sigma _{2}$.

Replacing global supersymmetry by canonical supergravity and setting $\phi
_{1}=\phi _{2}=0$ the potential $V(\sigma ,\sigma _{1},\sigma _{2})$ becomes 
\begin{equation}
V(\sigma ,\sigma _{1},\sigma _{2})=\mu ^{4}[(1-\frac{\sigma ^{2}}{2}+\frac{%
\sigma ^{4}}{4})+\frac{1}{4}\sigma ^{2}(\sigma _{1}^{2}+\sigma _{2}^{2})]%
\displaystyle{e^{\frac{1}{2}(\sigma ^{2}+\sigma _{1}^{2}+\sigma _{2}^{2})}}.
\end{equation}
This potential has now for fixed $\sigma >\sigma _{c}$ a minimum with
respect to $\sigma _{1}$ and $\sigma _{2}$ at $\sigma _{1}=\sigma _{2}=0$
with both fields acquiring at this minimum masses larger than $H$. Moreover,
setting $\sigma _{1}=\sigma _{2}=0$ in $V(\sigma ,\sigma _{1},\sigma _{2})$
and keeping $\sigma >\sigma _{c}$, we recover $V(\sigma )$ of eq. (6).
Supergravity, by giving large masses to $\sigma _{1}$ and $\sigma _{2}$,
chooses $\sigma $ as the inflaton and generates a potential for it.

The above heuristic arguments are only meant to illustrate the salient
features of the complete supergravity potential $V$ derived from $W$.
Careful analysis of this potential reveals that it possesses an absolute
minimum with respect to all fields at $\mid \frac{\phi_1}{2} \mid = \mid 
\frac {\phi_2}{2} \mid = \mu / \sqrt{\lambda_1 + \lambda_2}$ and $\sigma =
\sigma_1 =\sigma_2 =0$, where $V$ vanishes. Moreover, for fixed $%
\sigma>\sigma_c \simeq \sqrt{2} \mu / \sqrt{\lambda_1}$, $V$ has a minimum
with respect to the remaining fields at $\phi_1=\phi_2= \sigma_1=\sigma_2=0$%
. At this minimum the above fields have masses larger than $H$ and $%
V(\sigma) $ is given by eq. (6).

We now proceed to a more careful treatment of inflation with inflaton
potential $V(\sigma )$ given by eq. (6). Taking as a criterion for the
beginning of inflation the effective frequency $(\frac{V^{\prime }}{\sigma }%
)^{\frac{1}{2}}$ of $\sigma $ to be less than $\frac{3}{2}H$, we obtain $%
\frac{V^{\prime }}{V}\leq \frac{3}{4}\sigma $ or $\sigma ^{2}\leq \sqrt{61}%
-7\simeq 0.81$, i.e. $\sigma \leq 0.9$. Nevertheless, we only consider in
the following values of $\sigma \leq \frac{1}{\sqrt{2}}\simeq 0.7$. The
number of e-foldings in the slow roll approximation is given by $-\int \frac{%
V}{V^{\prime }}d\sigma $ (as an indefinite integral). We can take into
account, to a first approximation, the contribution of the second derivative
term and the kinetic energy of the inflaton by adding to this integral the
term $\frac{1}{3}ln\mid \frac{V^{\prime }}{V}\mid $. These effects are not
negligible for relatively large values of $\sigma $. The resulting function
of $\sigma $, which we denote by $N(\sigma )$, is then given by 
\begin{equation}
N(\sigma )\equiv -\int \frac{V}{V^{\prime }}d\sigma +\frac{1}{3}ln\mid \frac{%
V^{\prime }}{V}\mid \,=\sigma ^{-2}+\frac{3}{2}ln(\frac{\sigma ^{2}}{2})-%
\frac{5}{6}ln(1+\frac{\sigma ^{2}}{2})-\frac{1}{3}ln(1+\frac{\sigma ^{6}}{8}%
).
\end{equation}
The only parameter that $V(\sigma )$ depends on is the mass scale $\mu $ or
equivalently the false vacuum energy density $\mu ^{4}$. As a consequence
all quantities characterizing inflation, such as $\sigma _{H},\sigma _{c}$
and the (average) spectral index $n$, will depend on $\mu $ only. The
relation between $\mu $ and $\sigma _{H}$ is 
\begin{equation}
\mu ^{4}\simeq 720\pi ^{2}(\frac{\Delta T}{T})^{2}[(\frac{V}{V^{\prime }}%
)^{2}(\frac{V}{\mu ^{4}})+\frac{27}{4}]_{\sigma _{H}}^{-1},
\end{equation}
where the contribution of both the scalar and the tensor components to the
quadrupole anisotropy $\frac{\Delta T}{T}$ is taken into account. Let us
denote by $\ell _{H}$ the scale corresponding to our present horizon and by $%
\ell _{o}$ another length scale. Also let $\sigma _{o}$ be the value that
the inflaton field had when $\ell _{o}$ crossed outside the inflationary
horizon. We define the average spectral index $n(\ell _{o})$ for scales from 
$\ell _{o}$ to $\ell _{H}$ as 
\begin{equation}
n(\ell _{o})\equiv 1+2ln[(\frac{\delta \rho }{\rho })_{\ell _{o}}/(\frac{%
\delta \rho }{\rho })_{\ell _{H}}]/ln(\frac{\ell _{H}}{\ell _{o}})=1+2ln[(%
\frac{V^{3/2}}{V^{\prime }})_{\sigma _{o}}/(\frac{V^{3/2}}{V^{\prime }}%
)_{\sigma _{H}}]/\Delta N(\sigma _{H},\sigma _{o}).
\end{equation}
Here $(\delta \rho /\rho )_{\ell }$ is the amplitude of the energy density
fluctuations on the length scale $\ell $ as this scale crosses inside the
postinflationary horizon and $\Delta N(\sigma _{H},\sigma _{o})=N(\sigma
_{o})-N(\sigma _{H})=ln(\ell _{H}/\ell _{o})$. Finally, the value of $\sigma
_{c}$ is determined by requiring that $N_{H}\equiv \Delta N(\sigma
_{H},\sigma _{c})=N(\sigma _{c})-N(\sigma _{H})$ for some chosen value of $%
N_{H}$.

Table 1 gives the values of $\sigma _{H},\sigma _{c},$ $n\equiv n(\ell _{1})$
and $n_{COBE}\equiv n(\ell _{2})$, where $\ell _{1}$ $(\ell _{2})$ is the
scale that corresponds to $1Mpc$ $(2000Mpc)$ today, for different values of $%
\mu $ assuming that the present horizon size is $12000Mpc,\frac{\Delta T}{T}%
=6.6\times 10^{-6}$ and $N_{H}=50$. We see that canonical supergravity leads
naturally to very high values of the average spectral index which are higher
and more rapidly varying as the false vacuum energy density becomes higher.
Another distinctive feature of supergravity dominated hybrid inflation is
that the total number of e-foldings $\sim \sigma _{c}^{-2}$ is very limited
compared to most other inflationary scenarios.

When $\sigma $ falls below the critical value $\sigma _{c}\simeq \sqrt{2}\mu
/\sqrt{\lambda _{1}}$ the mass - squared term of $\phi _{1}$ becomes
negative and the local minimum of the potential at $\phi _{1}=\phi
_{2}=\sigma _{1}=\sigma _{2}=0$ becomes unstable. The subsequent evolution
is quite involved and can only be studied numerically. With a starting value
for $\mid \phi _{1}\mid $ of the order of a quantum fluctuation in de-Sitter
space $\mid \phi _{1}\mid $ becomes $\sim 2\mu /\sqrt{\lambda _{1}}$ in a
few Hubble times and inflation as described so far is terminated. At the
same time $\sigma _{1}$ grows. In order for $\mid \phi _{2}\mid $ to grow as
well the coupling $\lambda _{1}$ has to be a few times larger than $\lambda
_{3}$. Then, $\mid \phi _{2}\mid $ grows and gets stabilized at $\mid \phi
_{2}\mid \simeq 2\mu /\sqrt{\lambda _{2}}$. At the same time $\mid \phi
_{1}\mid $ leaves the value $2\mu /\sqrt{\lambda _{1}}$ and $\phi _{1}$
starts oscillating again around zero but now with a much larger amplitude.
The same happens with $\sigma _{1}$. Finally, after the energy density falls
a few orders of magnitude, $\mid \phi _{1}\mid $ gets stabilized at $\mid
\phi _{1}\mid =\mid \phi _{2}\mid =2\mu /\sqrt{\lambda _{1}+\lambda _{2}}$
with $\sigma ,\sigma _{1},\sigma _{2}$ oscillating around zero.

In the above discussion we insisted that the vevs of $\phi _{1}$ and $\phi
_{2}$ be equal and have the value suggested by the gauge coupling
unification of MSSM. This restriction can be easily relaxed by choosing the
value of $\frac{\lambda _{4}}{\lambda _{3}}=\frac{<\phi _{1}^{2}>}{<\phi
_{2}^{2}>}$. Then, $<\phi _{2}^{2}>=4\mu ^{2}/\lambda _{2}(1+\frac{\lambda
_{1}}{\lambda _{2}}\frac{\lambda _{4}}{\lambda _{3}})$ and $\sigma
_{c}^{2}\simeq 2\mu ^{2}/\lambda _{1}$, always$~$assuming that$~\lambda
_{2}>>\lambda _{1}$. We also insisted that the second non-inflaton field be
a gauge non-singlet. This restriction could be relaxed as well in order to
allow a non-inflaton field whose large vev breaks a continuous global
symmetry.

As a byproduct of our analysis it should be obvious that if in the original
non-supersymmetric hybrid inflationary model, where $\sigma _{c}$ is much
less constrained, one replaces the mass-squared term of the inflaton with a
suitably small quartic coupling one could easily obtain very large but
varying values of the spectral index without having to assume an enormous
vev for the non-inflaton field.

Before concluding we would like to point out briefly that our mechanism for
raising the value of $\sigma _{c}$ relative to the symmetry breaking scale
could be easily modified to lead to a value of $\sigma _{c}$ much lower than
the vevs of the non-inflaton fields. If we choose $\lambda _{1}$ in $W$ of
eq. (14) to be negative and $\lambda _{1}+\lambda _{2}<<-\lambda
_{1}<\lambda _{2}$, $\lambda _{3}=\lambda _{4}$, the resulting theory has a
SUSY vacuum at $\,\langle S\rangle =\langle S^{\prime }\rangle =0$,$\quad
<\Phi _{1}><\bar{\Phi}_{1}>\,=\,<\Phi _{2}><\bar{\Phi}_{2}>\,=$\thinspace $%
\mu ^{2}/(\lambda _{1}+\lambda _{2})$,\quad $\mid <\Phi _{1}>\mid $%
\thinspace $=$\thinspace $\mid <\bar{\Phi}_{1}>\mid \quad $and\quad $\mid
<\Phi _{2}>\mid $\thinspace $=$\thinspace $\mid <\bar{\Phi}_{2}>\mid $,$\ $%
with $\sigma _{^{_{_{c}}}}^{2}\simeq 2\mu ^{2}/\mid \lambda _{1}\mid $. This
observation could find applications in other occasions. For example, it
could be used in the scenario of ref. \cite{dvali} to raise the value of the
symmetry breaking scale to the MSSM value $M_{X}\simeq 2\times 10^{16}GeV$.

We conclude by summarizing our results. We investigated the possibility of
using as inflaton potential in SUSY hybrid inflation the one generated when
global supersymmetry is replaced by canonical supergravity. We argued that
in the context of MSSM this possibility is strongly connected with the
relaxation of the very tight relationship between the vev of the
non-inflaton field and the critical value of the inflaton field encountered
in the simplest SUSY models. We then presented a mechanism involving two
non-inflaton fields which accomplishes the relaxation of the unwanted
relationship. Finally, we studied the resulting supergravity dominated
inflationary scenario and pointed out that it leads naturally to blue
primordial spectra for adiabatic density perturbations.

\acknowledgments

This research was supported in part by EU under TMR contract
ERBFMRX-CT96-0090. The author would like to thank G. Lazarides for useful
discussions.

\newpage

\begin{tabular}{ccccc}
$\mu /10^{15}GeV$ & $\sigma _{H}/10^{17}GeV$ & $\sigma _{c}/10^{17}GeV$ & $%
\qquad n\qquad $ & $n_{COBE}$ \\ 
&  &  &  &  \\ 
1 & 2.0562 & 1.7634 & 1.021 & 1.022 \\ 
2 & 3.2524 & 2.3540 & 1.051 & 1.055 \\ 
4 & 5.1188 & 2.8232 & 1.119 & 1.139 \\ 
6 & 6.6435 & 3.0025 & 1.187 & 1.242 \\ 
8 & 7.9684 & 3.0894 & 1.252 & 1.358 \\ 
10 & 9.1556 & 3.1379 & 1.312 & 1.484 \\ 
12 & 10.2415 & 3.1678 & 1.367 & 1.618 \\ 
15 & 11.7342 & 3.1945 & 1.442 & 1.830 \\ 
20 & 14.0129 & 3.2171 & 1.553 & 2.209
\end{tabular}
\newline
\newline

\noindent \quad Table 1. The values of $\sigma _{H}$, $\sigma _{c}$, $n$ and 
$n_{COBE}$ as a function of $\mu $.

\end{document}